\documentclass[prd,aps,showpacs,tightenlines]{revtex4}  
\usepackage{mathrsfs}
\usepackage{amsmath}
\usepackage{amssymb}
\usepackage{epsfig}
\usepackage{graphicx}
\usepackage{booktabs}
\usepackage{multirow}
\usepackage{subfigure}

\begin{document}
\newcommand{\psl}{ p \hspace{-1.8truemm}/ }
\newcommand{\nsl}{ n \hspace{-2.2truemm}/ }
\newcommand{\vsl}{ v \hspace{-2.2truemm}/ }
\newcommand{\epsl}{\epsilon \hspace{-1.8truemm}/\,  }

\title{Revisiting   nonfactorizable contributions to  factorization-forbidden decays of $B$ mesons to charmonium }
\author{Ya-Qian Li$^1$}
\author{Meng-Kun Jia$^1$}
\author{Zhou Rui$^1$}\email[Corresponding  author: ]{jindui1127@126.com}
\affiliation{$^1$College of Sciences, North China University of Science and Technology,
                          Tangshan 063009,  China}
\date{\today}
\begin{abstract}
Motivated by the large rates of $B\rightarrow (\chi_{c0}, \chi_{c2}, h_c)K$ decays observed by the $BABAR$ and Belle collaborations,
we investigate the nonfactorizable contributions to these factorization-forbidden decays,
which can occur  through  a gluon exchange between the $c\bar c$ system and  the spectator quark.
Our numerical results demonstrate that the spectator contributions are capable
of producing a large branching ratio consistent with the experiments.
As a by-product, we also study the Cabibbo-suppressed decays,
such as $B\rightarrow (\chi_{c0}, \chi_{c2}, h_c)\pi$ and
the U-spin-related $B_s$ decay,
which have so far received less theoretical and experimental attention.
The calculated branching ratios reach the order of $10^{-6}$,
which  in within the scope of the Belle-II and LHCb experiments.
Further, the $CP$-asymmetry parameters are also calculated for these decays.
The obtained results are compared with the available experimental data and numbers from other predictions.
We also investigate  the sources of theoretical uncertainties in our calculation.
\end{abstract}



\maketitle

\section{Introduction}
In the Standard Model (SM) of particle physics, the charmonium decays of $B$ meson
 arise from the quark-level process $b\rightarrow q c\bar{c}$ with $q=d,s$,
involving tree and penguin amplitudes.
The $S$-wave charmonium states are produced from a $c\bar c$
system with the orbital angular momentum $L=0$, such as $\eta_c$ and $J/\psi$.
These decays belong to the color-suppressed category
and receive large nonfactorizable contributions.
For the orbital excitation of the $c\bar c$ assignments with  $L=1$,
the spin and orbit interaction between the charm-anticharm quarks pair can
create four $P$-wave charmonium states, namely, $\chi_{c0}$, $\chi_{c1}$, $\chi_{c2}$, and $h_c$.
Except for the $\chi_{c1}$ modes, which are allowed under the factorization hypothesis,
other modes are prohibited  because of  the $V-A$ structure of the weak vertex~\cite{plb59191,jhep06067},
where $V$ and $A$ denote the vector and axial vector currents, respectively.
However, these processes  can occur through a gluon exchange between the charmonium system and the quark in other mesons,
which induces the so-called nonfactorizable contributions~\cite{jhep06067}.
Therefore, the factorization-forbidden  decays of $B$ meson to a $P$-wave charmonium state
can provide valuable insights into the nonfactorizable mechanism.

Experimentally, the first observation of the decay $B^+ \rightarrow \chi_{c0} K^+$ was
reported by the Belle Collaboration~\cite{prl88031802} using $\chi_{c0}$ decays to the pion or kaon pair,
later confirmed by the  $BABAR$ Collaboration~\cite{prd69071103}.
Subsequently, both the  $BABAR$~\cite{prd74071101,prd78012004,prd84092007,prd85112010}
and Belle Collaborations~\cite{prd71092003,prl96251803}  performed improved measurements.
The current world averages of the absolute branching ratio $B \rightarrow \chi_{c0} K$ have reached the order of $10^{-4}$~\cite{pdg2018},
which is of the same order of magnitude as that of the factorization-allowed $\chi_{c1}$ mode.
The corresponding vector $K^*$ mode has been searched for and observed
by the $BABAR$ collaboration~\cite{prl94171801,prd78091101} with a similar rate.
Another factorization-inhibited decay, $B^+\rightarrow \chi_{c2}K^+$, has been measured by the Belle~\cite{plb634155,prl107091803}
and $BABAR$~\cite{prl102132001} collaborations with an average branching ratio of $(1.1\pm0.4)\times 10^{-5}$~\cite{pdg2018},
 which is  an order of magnitude smaller than that of the $\chi_{c0}$ mode.
Several collaborations have also searched for the process $B\rightarrow h_c K$~\cite{prd74012007,prd78012006,epjc732462}.
The upper limit of the current branching ratio  is $3.8\times 10^{-5}$
at $90\%$ confidence level~\cite{pdg2018}, obtained during the search for $h_c$ by the Belle collaboration~\cite{prd74012007}.
Very recently, the Belle Collaboration performed a search for the decays $B^+\rightarrow h_c K^+$ and $B^0\rightarrow h_c K^0_S$~\cite{prd100012001}.
They found evidence for the former with a $4.8\sigma$ significance  and set an  upper limit on the latter.
These results are comparable to $B\rightarrow \chi_{c2} K$  but below the measured rate for $B\rightarrow \chi_{c0} K$.
Most recently, both the Belle~\cite{prd97012005} and $BABAR$~\cite{prl124152001} Collaborations present the measurement of the absolute branching fractions of $B^+\rightarrow X_{cc} K^+$, where $X_{cc}$  refers to a charmonium state.

The measured branching ratios in these factorization-inhibited decays are surprisingly
larger than the expectations from factorization,
which suggests that the nonfactorizable contributions in $B$ decays to charmonium can be sizable.
Several attempts have been made to address these challenges.
Meli$\acute{c}$ analyzed the   effects of soft nonfactorization on
$B\rightarrow (\eta_c,J/\psi,\chi_{c0,c1})K$ decays~\cite{plb59191}  using the light-cone sum rules (LCSR) approach.
The calculated branching ratio of the reaction  $B\rightarrow \chi_{c0}K$, however, is too small to accommodate the data
because of  the larger cancellation between the twist-3 and twist-4 pieces in the nonfactorizable contributions.
A similar conclusion is also drawn in~\cite{prd70074006} using the same method.
The exclusive $B$ decays to the $P$-wave charmonium states   have been  studied within the framework of QCD factorization
(QCDF)~\cite{plb568127,prd69054009,plb619313,ctp48885,0607221,prd87074035}.
It was  found that the soft contributions may be large as infrared divergences arise from
vertex corrections as well as endpoint singularities owing to the leading twist spectator corrections.
 However, Beneke pointed out that the concerned decays  may occur through the color octet mechanism~\cite{prd59054003},
and  the endpoint divergence in hard spectator scattering factorizes and can be  absorbed into color octet operator matrix elements~\cite{npb811155}.
The $B\rightarrow \chi_{c0} K$ and $B\rightarrow h_c K$ decays~\cite{prd71114008,prd74114029} have been investigated under the perturbative QCD (PQCD) approach based on the $k_T$ factorization theorem~\cite{pqcd1,pqcd2,pqcd3,pqcd4},
in which the vertex corrections are ignored and the endpoint singularity is cured by including the parton's transverse momentum.
Without endpoint singularity, the PQCD approach has so far been successfully applied to the
studies of various factorization-allowed charmonium decays of the $B$ meson
\cite{jhep03009,cpc34937,cpc341680,prd86011501,prd89094010,plb772719,cpc11113102,epjc77610,epjc77199,prd97033006,prd98113003,prd99093007,prd91094024}.
Most predictions are consistent with the current experiments.
Under another approach~\cite{prd69054023,plb54271}, by considering intermediate charmed meson rescattering effects,
the authors claimed that the rescattering effects could provide the large part of the $B\rightarrow \chi_{c0}K, h_c K$ amplitudes.

Generally, the nonfactorizable dynamics include both the vertex corrections and spectator amplitudes.
We agree with the comment made in \cite{prd71114008}:
the PQCD formalism for the vertex corrections requires
the charmonium meson wave functions, dependent on the transverse momenta,
 as the necessary nonperturbative inputs,
which are not yet available completely.
The authors of~\cite{ctp48885}  confirm that the vertex corrections are numerically small,
and the spectator corrections are large and dominant in the QCDF.
Furthermore, the previous PQCD calculations~\cite{prd71114008,prd74114029}
demonstrate that the spectator contributions are sufficient to account for the measurements.
Motivated by the same idea, we have reason to believe that it is appropriate to analyze other factorization-inhibited decays,
such as $B\rightarrow \chi_{c2} K$, in the PQCD approach.
Moreover, we update the $P$-wave charmonium distribution amplitudes (DAs) based on our previous study~\cite{prd97033001},
 where the new universal nonperturbative objects are successful
 in describing the $P$-wave charmonium decays of the $B/B_c$ meson \cite{prd97033001,prd98033007,epjc78463}.

The layout of this paper is as follows: In Section~\ref{sec:framework},
we present our theoretical formulae based on the PQCD framework.
The input parameters together with the numerical results and discussions are presented in Section~\ref{sec:results}.
Finally, we present the  summary in Section~\ref{sec:sum}.
The relevant meson distribution amplitudes are presented in the Appendix.

\section{ ANALYTIC FORMULAS}\label{sec:framework}
\begin{figure}[tbp]
\centerline{\epsfxsize=9cm \epsffile{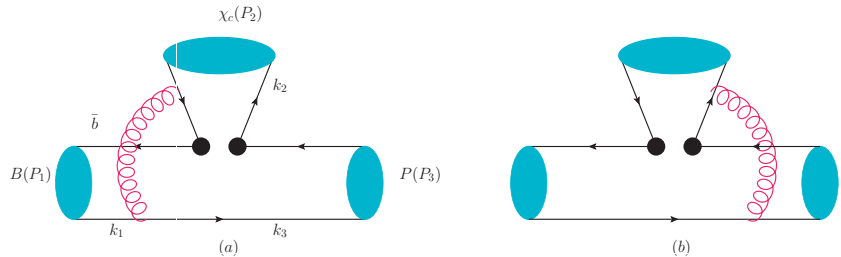}}
\vspace{1cm}
\vspace{-8cm}
\caption{Nonfactorizable spectator diagrams for  decays  $B\to \chi _{c} P$,
where $P$ stands for  final-state hadron pion or kaon.}
\label{fig:femy}
\end{figure}
For the factorization-inhibited  $B$ decays in question,
only the nonfactorization spectator diagrams contribute, which are displayed in Fig~\ref{fig:femy}.
In the rest frame of the $B$ meson, we define $P_i$ and $k_i$ for $i=1,2,3$ to be the four-momenta
of the mesons and quarks in the initial and final states, as indicated in Fig~\ref{fig:femy}.
In terms of the light-cone coordinates, they are parametrized as~\cite{epjc78463}
\begin{eqnarray}
 P_1&=&\frac{M}{\sqrt{2}}(1,1,\textbf{0}_{\rm T}),\quad P_2=\frac{M}{\sqrt{2}}(1,r^2,\textbf{0}_{\rm T}),\quad  P_3=\frac{M}{\sqrt{2}}(0,1-r^2,\textbf{0}_{\rm T}),\nonumber\\
  k_1&=&(\frac{M}{\sqrt{2}}x_1,0,\textbf{k}_{\rm 1T}),\quad k_2=(\frac{M}{\sqrt{2}}x_2,\frac{M}{\sqrt{2}}x_2r^2,\textbf{k}_{2\rm T}),\quad  k_3=(0,\frac{M}{\sqrt{2}}x_3(1-r^2),\textbf{k}_{\rm 3T}),
\end{eqnarray}
where $r=m/M$ represents the ratio of the charmonium  mass $m$ to the $B$ meson mass $M$.
We  neglect the light pseudoscalar meson mass for simplicity.
 $x_i$ and $k_{iT}$  represent the parton longitudinal momentum fractions and parton transverse momenta, respectively.
Similar to  the vector charmonium,
the longitudinal  polarization vectors $\epsilon_L$ of an axial-vector charmonium can be defined as
\begin{eqnarray}\label{eq:polar}
\epsilon_{L}&=&\frac{1}{\sqrt{2}r}(1,-r^2,\textbf{0}_{\rm T}),
\end{eqnarray}
which satisfy the normalization $\epsilon^2_{L}= -1$
and the orthogonality $\epsilon_{L} \cdot P_2=0$.
For a tensor charmonium, the polarization tensor $\epsilon_{\mu\nu}(\lambda)$ with
helicity $\lambda$ can be constructed via the polarization vector $\epsilon_{\mu}$~\cite{prd82054019,prd83034001,prd83014008},
whose detail expressions can be found in Refs~\cite{prd97033001,prd98033007}.

The Hamiltonian referred to in the SM is written as~\cite{rmp681125}:
\begin{eqnarray}\label{eq:operator}
\mathcal{H}_{eff}=\frac{G_F}{\sqrt{2}}\{\xi_c[C_1(\mu)O^c_1(\mu)+C_2(\mu)O^c_2(\mu)]
-\xi_t\sum_{i=3}^{10}C_i(\mu)O_i(\mu)\},
\end{eqnarray}
where $G_F$ is the Fermi coupling constant and  $\xi_{c(t)}=V^*_{c(t)b}V_{c(t)q}$
is the production of the Cabibbo-Kobayashi-Maskawa (CKM) matrix element.
$O_i(\mu)$ and $C_i(\mu)$ are the local four-quark operators and
their QCD-corrected Wilson coefficients at the renormalization scale $\mu$, respectively.
Their explicit expressions can be found in Ref.~\cite{rmp681125}.

In the PQCD approach, the decay amplitudes are expressed
as the convolution of the hard kernel $H$ with the relevant
meson light-cone wave functions $\Phi_i$
\begin{eqnarray}
\label{eq:ampu}
\mathcal{A}(B\rightarrow \chi_c P) = \int d^4k_1d^4k_2d^4k_3 Tr[C(t)\Phi_B(k_1)\Phi_{\chi_c}(k_2)\Phi_P(k_3)
H(k_1,k_2,k_3,t)],
\end{eqnarray}
where ``Tr'' denotes the trace over all Dirac structures and color indices. $t$ is the energy scale in the hard function $H$.
The meson wave functions $\Phi$ absorb the nonperturbative dynamics in the hadronization processes.
For  definitions of the $P$-wave charmonium  wave functions, please refer to our  previous study~\cite{prd97033001},
whereas for those  $B$ and light pseudoscalar mesons, one can consult Ref.~\cite{prd76074018}.
We list the relevant meson distribution amplitudes  and the corresponding parameters in the Appendix.
As mentioned before, because only the nonfactorizable spectator diagrams contribute to the hard kernel $H$,
the momentum convolution integral in Eq.~(\ref{eq:ampu}) would involve all the initial and final states altogether.
The perturbative calculations can be performed  without endpoint singularity.
In the following, we  compute the  amplitudes of the concerned decays.

We mark subscripts $S$, $A$, and $T$ to denote the scalar, axial-vector, and tensor charmonium
in the final states, respectively.
The decay amplitudes for $(V-A)\otimes(V-A)$ operators  read as
\begin{eqnarray}\label{eq:mllls}
\mathcal{M}_{S}&=&-16\sqrt{\frac{2}{3}}\pi C_f M^4\int_0^1dx_1dx_2dx_3\int_0^{\infty}b_1b_2db_1db_2 \phi_B(x_1,b_1)\nonumber\\&&
\{-\psi_S^s(x_2)2r_cr[\phi^A_P(x_3)-4r_p\phi^P_P(x_3)]+\nonumber\\&&\psi_S^v(x_2)[2r_p\phi^P_P(x_3)(r^2(x_1+x_3-2x_2)-x_3)
+\phi^A_P(x_3)(r^2(x_1-2x_3)-2x_1+2x_2+x_3) ]\}
\nonumber\\&&\alpha_s(t)S_{cd}(t)h(\alpha,\beta,b_1,b_2),
\end{eqnarray}
\begin{eqnarray}\label{eq:mlllA}
\mathcal{M}_{A}&=&-16\sqrt{\frac{2}{3}}\pi C_f M^4\int_0^1dx_1dx_2dx_3\int_0^{\infty}b_1b_2db_1db_2 \phi_B(x_1,b_1)\nonumber\\&&
\{\psi_A^L(x_2)[2r_p\phi^P_P(x_3)(r^2(x_1-x_3)+x_3)-\phi^A_P(x_3)(r^2(3x_1-2x_2-2x_3)-2x_1+2x_2+x_3)]\nonumber\\&&
-4rr_cr_p\psi_A^t(x_2)\phi^T_P(x_3)\}\alpha_s(t)S_{cd}(t)h(\alpha,\beta,b_1,b_2),
\end{eqnarray}
\begin{eqnarray}\label{eq:mlllT}
\mathcal{M}_{T}&=&-\frac{32}{3}\pi C_f M^4\int_0^1dx_1dx_2dx_3\int_0^{\infty}b_1b_2db_1db_2 \phi_B(x_1,b_1)\nonumber\\&&
\{\psi_T(x_2)[2r_p\phi^P_P(x_3)(r^2(x_1-x_3)+x_3)-\phi^A_P(x_3)(r^2(3x_1-2x_2-2x_3)-2x_1+2x_2+x_3)]\nonumber\\&&
-4rr_cr_p\psi_T^t(x_2)\phi^T_P(x_3)\}\alpha_s(t)S_{cd}(t)h(\alpha,\beta,b_1,b_2),
\end{eqnarray}
 with the color factor $C_f=4/3$ and the mass ratios $r_{c}=m_{c}/M$ and  $r_p=m_0/M$, where $m_c$ is the charm quark mass,
and  $m_0$ is the chiral scale parameter.
For simplicity, in the above formulas, we have combined the two decay amplitudes in Fig.~\ref{fig:femy}(a) and \ref{fig:femy}(b),
because their hard kernels are symmetric under $x_2 \leftrightarrow 1-x_2$,
and dropped the power-suppressed terms higher than $r^2$.
$\alpha$ and $\beta$ represent the virtuality of the internal gluon and quark, respectively,
expressed as
\begin{eqnarray}\label{eq:mm}
\alpha =x_1x_3(1-r^2)M^2, \quad \beta=[(x_1-x_2)(x_3+r^2(x_2-x_3))+r_c^2]M^2.
\end{eqnarray}
The hard scale $t$ is chosen as the largest scale of the virtualities of the internal particles
in the hard amplitudes:
\begin{eqnarray}
 \quad t=\max(\sqrt{\alpha},\sqrt{\beta},1/b_1,1/b_2).
\end{eqnarray}
The details of the functions $h$ and the Sudakov factors $S_{cd}(t)$
are provided in Appendix A of Ref.~\cite{epjc77610}.
To determine  the decay amplitudes from the $(V-A)\otimes(V+A)$ operators,
we carry out the  Fierz transformation to obtained the appropriate  color structure for factorization to work;
they satisfy the relations
\begin{eqnarray}
\mathcal{M'}_{S,T}=\mathcal{M}_{S,T},\quad \mathcal{M'}_{A}=-\mathcal{M}_{A}.
\end{eqnarray}

By combining the contributions from different diagrams with the corresponding Wilson coefficients,
 we obtain  the total decay amplitudes as
\begin{eqnarray}\label{eq:alnt}
\mathcal{A}&=&\xi_c C_2\mathcal{M} -\xi_t\Big [(C_4+C_{10})\mathcal{M}+(C_6+C_8)\mathcal{M'}\Big ],
\end{eqnarray}
and the $CP$-averaged branching ratio is then expressed as
\begin{eqnarray}
\mathcal {B}(B\rightarrow \chi_{c}P)=\frac{G_F^2\tau_{B}}{32\pi M}(1-r^2)
\frac{|\mathcal {A}|^2+|\mathcal {\bar{A}}|^2}{2},
\end{eqnarray}
where $\mathcal {\bar{A}}$ denotes the corresponding charge conjugate decay amplitude,
which can be obtained by conjugating the CKM elements in $\mathcal {A}$.
\section{Numerical results}\label{sec:results}
To estimate the contributions from the decay amplitudes,
we must specify the various parameters using in  our numerical analysis throughout this paper.
We employ  the meson lifetimes $\tau_{B^+}=1.638$ ps,  $\tau_{B_s}=1.51$ ps, and $\tau_{B^0}=1.51$ ps \cite{pdg2018}.
The parameters relevant for the Wolfenstein parameters are
$\lambda = 0.22506$, $A=0.811$,  $\bar{\rho}=0.124$, and $\bar{\eta}=0.356$ \cite{pdg2018}.
We consider $M_B=5.28$ GeV,  $M_{B_s}=5.37$ GeV, $m_{\chi_{c0}}=3.415$ GeV,
 $m_{\chi_{c2}}=3.556$ GeV,  and $m_{h_c}=3.525$ GeV for the meson masses \cite{pdg2018} and
 $\bar{m}_b(\bar{m}_b)=4.18$ GeV, $\bar{m}_c(\bar{m}_c)=1.275$ GeV
 for the $b$ and $c$ quark ``running" masses in the $\overline{MS}$ scheme.
The chiral masses  relate the pseudoscalar meson mass to the quark mass, which is  set as $m_0=1.6\pm 0.2$ GeV \cite{jhep01010}.
The decay constants (GeV) can be extracted from other decay rates or evaluated from the QCD sum rules, which are summarized here \cite{pdg2018,prd76074018,prd96014026}
\begin{eqnarray}\label{eq:con}
f_B&=&0.19\pm 0.02, \quad  f_{B_s}=0.23\pm0.02, \quad  f_{\pi}=0.131, \quad  f_{K}=0.16, \nonumber\\
f_{\chi_{c0}}&=&0.36,\quad  f_{\chi_{c2}}=0.177, \quad f_{\chi_{c2}}^{\perp}=0.128,\quad
f_{h_c}=0.127,\quad  f_{h_c}^{\perp}=0.133.
\end{eqnarray}

Using these input parameters and employing the analytic formulas presented in Section~\ref{sec:framework},
we derive the $CP$-averaged branching ratios with error bars as follows:
 \begin{eqnarray}\label{eq:brs}
\mathcal {B}(B^+\rightarrow \chi_{c0}K^+)&=&(1.4^{+0.4+0.3+0.7+0.5+0.7}_{-0.3-0.3-0.6-0.3-0.4})
\times 10^{-4}=(1.4^{+1.3}_{-0.9})\times 10^{-4},\nonumber\\
\mathcal {B}(B^+\rightarrow \chi_{c2}K^+)&=&(3.2^{+0.8+0.6+0.4+0.2+0.9}_{-0.7-0.7-0.5-0.4-0.6})
\times 10^{-5}=(3.2^{+1.4}_{-1.3})\times 10^{-5},\nonumber\\
\mathcal {B}(B^+\rightarrow h_{c}K^+)&=&(3.5^{+1.0+0.7+0.5+0.2+1.3}_{-0.8-0.7-0.4-0.2-0.7})
\times 10^{-5}=(3.5^{+1.9}_{-1.3})\times 10^{-5},\nonumber\\
\mathcal {B}(B^+\rightarrow \chi_{c0}\pi^+)&=&(3.6^{+1.0+0.8+2.4+1.5+2.1}_{-0.7-0.8-1.6-0.8-1.3})
\times 10^{-6}=(3.6^{+3.7}_{-2.4})\times 10^{-6},\nonumber\\
\mathcal {B}(B^+\rightarrow \chi_{c2}\pi^+)&=&(1.0^{+0.3+0.2+0.2+0.1+0.3}_{-0.2-0.2-0.2-0.1-0.2})
\times 10^{-6}=(1.0^{+0.5}_{-0.4})\times 10^{-6},\nonumber\\
\mathcal {B}(B^+\rightarrow h_{c}\pi^+)&=&(1.1^{+0.3+0.2+0.2+0.1+0.4}_{-0.2-0.2-0.2-0.1-0.2})
\times 10^{-6}=(1.1^{+0.6}_{-0.4})\times 10^{-6},\nonumber\\
\mathcal {B}(B_s\rightarrow \chi_{c0}\bar{K}^{0})&=&(4.3^{+1.4+0.8+2.7+1.8+2.4}_{-1.0-0.7-2.0-1.1-1.6})
\times 10^{-6}=(4.3^{+4.4}_{-3.0})\times 10^{-6},\nonumber\\
\mathcal {B}(B_s\rightarrow \chi_{c2}\bar{K}^{0})&=&(1.1^{+0.4+0.2+0.1+0.2+0.5}_{-0.2-0.2-0.3-0.1-0.2})
\times 10^{-6}=(1.1^{+0.7}_{-0.5})\times 10^{-6},\nonumber\\
\mathcal {B}(B_s\rightarrow h_c\bar{K}^{0})&=&(1.2^{+0.4+0.2+0.2+0.1+0.5}_{-0.3-0.2-0.2-0.1-0.3})
\times 10^{-6}=(1.1^{+0.7}_{-0.5})\times 10^{-6},
\end{eqnarray}
where the second ``equal to" sign in each row denote the central
value with all uncertainties added in quadrature.
The theoretical errors correspond to the uncertainties owing to
the shape parameters (1) $\omega_b=0.40\pm 0.04$ GeV for the $B$ meson and $\omega_b=0.50\pm 0.05$ GeV for the $B_s$ meson,
(2)  decay constant of $B/B_s$ meson, which is expressed in Eq.~(\ref{eq:con}),
(3)  chiral scale parameter $m_0=1.6\pm 0.2$ GeV \cite{jhep01010} associated with the kaon or pion,
which reflects the uncertainty in the current quark masses,
(4) heavy quark masses $m_{c(b)}$ within a $20\%$ range,
and (5) hard scale $t$, defined in Eqs.~(\ref{eq:mllls})-(\ref{eq:mlllT}),
 which we vary from $0.75t$ to $1.25t$, and $\Lambda^{(5)}_{QCD}=0.25\pm 0.05$ GeV.
In general, the uncertainties induced by these parameters  are comparable.
The uncertainties stemming from the decay constants of charmonium states are not presented  explicitly,
which affect on the branching ratios via the relation   $\mathcal{B}\propto f_{\chi_c}^2$.
We have matched the sensitivity of the branching ratios to the  charm quark velocity $v$
inside the  charmonium DAs in Eq.~(\ref{eq:vv2}).
The variation of $v^2$ in the range 0.25-0.35
indicates that  the difference in our results does not exceed 10\%,
which suggests that the relativistic corrections
 may  not be significant for these decays.

Further, because isospin is conserved in the heavy quark limit,
we can obtain the branching ratios of the neutral counterpart
by multiplying the charged ones by the lifetime ratio $\tau_{B^0}/\tau_{B^+}$:
\begin{eqnarray}
\mathcal {B}(B^0\rightarrow \chi_{c}K^0)=\frac{\tau_{B^0}}{\tau_{B^+}}\mathcal {B}(B^+\rightarrow \chi_{c}K^+),\quad
\mathcal {B}(B^0\rightarrow \chi_{c}\pi^0)=\frac{\tau_{B^0}}{2\tau_{B^+}}\mathcal {B}(B^+\rightarrow \chi_{c}\pi^+).
\end{eqnarray}

From Eqs.~(\ref{eq:mllls})-(\ref{eq:mlllT}), it can be seen that each decay amplitude
receives contributions from both the twist-2 and twist-3  DAs of the charmonium state,
the results  of which are displayed separately in Table~\ref{tab:twist}, with all the input parameters considered at their central values.
We simply symbolize them as ``twist-2" and ``twist-3", respectively,
while the label ``total" corresponds to  the total contributions.
\begin{table}
\caption{Values of decay amplitudes from twist-2 and twist-3 charmonium distribution
amplitudes for $B^+\rightarrow (\chi_{c0},\chi_{c2},h_c)K^+$ decays.
Results are presented in units of $10^{-3}$GeV$^{3}$.}
\label{tab:twist}
\begin{tabular}[t]{lccc}
\hline\hline
Decay amplitudes  &Twist-2  & Twist-3  & Total  \\ \hline
$\mathcal {A}(B^+ \rightarrow \chi_{c0} K^+) $  &13+21i &-4-4i &9+17i\\
$\mathcal {A}(B^+ \rightarrow \chi_{c2} K^+) $   &5.9+8.3i &-0.2-0.6i &5.7+7.7i \\
$\mathcal {A}(B^+ \rightarrow h_c K^+) $   &5.9+8.5i &-1.1+0.2i &4.8+8.7i \\
\hline\hline
\end{tabular}
\end{table}
The numerical results  indicate that the dominant contribution comes from the twist-2 DA.
This can be understood from the formulas expressed in Eqs.~(\ref{eq:mllls})-(\ref{eq:mlllT}):
the contribution of the twist-3 DA is power-suppressed as compared to that of  the twist-2 DA.
As we have used the same asymptotic model of the  twist-2 DA for  the three charmonium states [see Eq.~(\ref{eq:cDAs})],
these decay amplitudes are governed by their different decay constants.
The relation among the decay constants $f_{\chi_{c0}}>f_{\chi_{c2}}\sim f_{h_c}$  expressed in Eq.~(\ref{eq:con}) roughly implies
the hierarchy pattern presented  in Table~\ref{tab:twist}.
For the twist-3 piece, the various asymptotic behaviors and tensor decay constants  contribute to different values.

As discussed in Refs~\cite{epjc50877,prd74114010},
for the nonleptonic two-body decay of the $B$ meson,
if the emitted particle from the weak vertex is a pseudoscalar or vector meson,
there is a destructive interference  between the two nonfactorizable spectator  diagrams presented in Fig.\ref{fig:femy}
owing to the symmetric twist-2 DAs of the emitted meson.
However, in  this study, all the twist-2 DAs of the emitted charmonium state
are  antisymmetric under the exchange of  the momentum fraction $x_2$ of the $c$ quark
and $1-x_2$ of the $\bar c$ quark [see Eq.~(\ref{eq:cDAs})],
which reverses the constructive or destructive interference situation.
To be more explicit, we present the values of  two spectator  amplitudes in Table~\ref{tab:cdtu},
where  $\mathcal {A}_a$  and $\mathcal {A}_b$ denote the decay amplitudes
from Fig.~\ref{fig:femy}(a) and Fig.~\ref{fig:femy}(b), respectively.
\begin{table}
\caption{Values of decay amplitudes from nonfactorizable diagrams (a) and (b)
for $B^+\rightarrow (\chi_{c0},\chi_{c2},h_c)K^+$ decays.
Results are presented in units of $10^{-3}$GeV$^{3}$.}
\label{tab:cdtu}
\begin{tabular}[t]{lccc}
\hline\hline
Decay amplitudes  & $\mathcal {A}_a$  & $\mathcal {A}_b$  & Total  \\ \hline
$\mathcal {A}(B^+ \rightarrow \chi_{c0} K^+) $  &3+3i &6+14i &9+17i\\
$\mathcal {A}(B^+ \rightarrow \chi_{c2} K^+) $   &3.3+0.4i &2.4+7.3i &5.7+7.7i \\
$\mathcal {A}(B^+ \rightarrow h_c K^+) $   &1.8+0.5i &3.0+8.2i &4.8+8.7i \\
\hline\hline
\end{tabular}
\end{table}
It can be seen that the constructive   interference between $\mathcal {A}_a$  and $\mathcal {A}_b$  can enhance the total decay amplitudes.
Therefore the large decay rates for these factorization-forbidden modes are comparable to those naively factorizable decays.
\begin{table}
\caption{Branching ratios (in units of $10^{-4}$) of  Cabibbo-favored decays from various theoretical studies in  literature
\cite{plb59191,prd70074006,npb811155,ctp48885,0607221,plb619313,prd71114008,prd74114029}. Data have been  taken from PDG 2018 \cite{pdg2018}. }
\label{tab:br}
\begin{tabular}[t]{lccccc}
\hline\hline
Modes  &This Work  & LCSR  & QCDF   &PQCD  &Data \cite{pdg2018} \\ \hline
$B^+ \rightarrow \chi_{c0} K^+ $  &$1.4^{+1.3}_{-0.9} $ & $(1.7\pm 0.2)\times 10^{-3}$ \cite{plb59191}
&$0.78^{+0.46}_{-0.35}$ \cite{plb619313} &5.61 \cite{prd71114008} &$1.50^{+0.15}_{-0.13}$ \\
&&  $1.0 \pm 0.6$ \cite{prd70074006}& $2 \sim 4$ \cite{ctp48885} &&\\
&&& 1.05 \cite{0607221} \footnotemark[2]&&\\
$B^0 \rightarrow \chi_{c0} K^0 $   & $1.3^{+1.2}_{-0.8}$ &-- & $1.13 \sim 5.19$ \cite{npb811155} \footnotemark[1]
& 5.24 \cite{prd71114008} &$1.11^{+0.24}_{-0.21}\times 10^{-2}$ \\
$B^+ \rightarrow \chi_{c2} K^+ $ & $0.32^{+0.14}_{-0.13}$ &--&$1.68^{+0.78}_{-0.69}$ \cite{plb619313}&--&$0.11\pm 0.04$\\
&&& 0.03 \cite{0607221} \footnotemark[2] &&\\
$B^0 \rightarrow \chi_{c2} K^0 $ & $0.29^{+0.13}_{-0.12}$ &--&$0.28 \sim 3.98$ \cite{npb811155} \footnotemark[1] &--&$<0.15$\\
$B^+ \rightarrow h_c K^+ $ & $0.35^{+0.19}_{-0.13}$ &--&0.27 \cite{0607221} \footnotemark[2] &0.36 \cite{prd74114029}&$0.37\pm 0.12$\\
$B^0 \rightarrow h_c K^0 $ & $0.32^{+0.18}_{-0.12}$ &--&$0.29 \sim 0.53$ \cite{npb811155} \footnotemark[1]&--&$<0.14$ \footnotemark[3]\\
\hline\hline
\end{tabular}
\footnotetext[1]{Quoted range represents   variation in charm quark mass.}
\footnotetext[2]{ We quote  result with $\mu=4.4$ GeV. }
\footnotetext[3]{Cited upper limit  $1.4\times 10^{-5}$ is for $B^0\rightarrow h_c K^0_S$ \cite{prd100012001}. }
\end{table}

To make a comparison, we also collect the various available theoretical predictions evaluated in LCSR \cite{plb59191,prd70074006},
QCDF \cite{npb811155,plb619313,ctp48885,0607221}, and PQCD \cite{prd71114008,prd74114029}
as well as the current world average values from the PDG \cite{pdg2018} presented in Table \ref{tab:br}.
The LCSR  calculations mainly focus on the $\mathcal {B}(B^+ \rightarrow \chi_{c0} K^+)$ decay.
There are lager discrepancies in their numerical results \cite{plb59191,prd70074006}.
The LCSR prediction of $\mathcal {B}(B^+ \rightarrow \chi_{c0} K^+)\sim 10^{-7}$ \cite{plb59191}
is far too small and clearly ruled out by experiment.
Three QCDF calculations exist for the concerned decays.
As mentioned in the Introduction, the QCDF suffers the infrared divergences arising from the vertex diagrams and
 endpoint  divergences owing to spectator amplitudes in the leading-twist order.
The main reason for  the different numerical results is the treatment of these divergences.
 In~\cite{ctp48885}, the infrared and endpoint divergences were regularized by a non-zero gluon mass and
the off-shellness of the quarks, respectively.
The authors  found that  the $B\rightarrow \chi_{c0} K$ is dominated by the  spectator contribution,
whereas  the contributions arising from the vertex are numerically small.
Subsequently, the same scheme is applied to the $B\rightarrow h_c(\chi_{c2}) K$ decays~\cite{0607221}
with the exception of neglecting the vertex corrections.
In Ref.~\cite{npb811155},  the authors used the  space-time dimension  as the  infrared regulator,
whereas the endpoint divergence was absorbed into the color octet operator matrix elements.
The scheme in~\cite{plb619313} is that a non-zero binding energy is introduced to regularize
the infrared divergence, and the spectator contributions are parameterized in a model-dependent way.
In general, our results are more consistent with the QCDF predictions from~\cite{npb811155}
within the low charm quark mass region.
Compared with previous PQCD calculations~\cite{prd71114008,prd74114029},
we update the charmonium distribution amplitudes and some input parameters in this study.
Our predictions on $\mathcal {B} (B\rightarrow \chi_0 K)$  yield typically smaller values than those of~\cite{prd71114008}
and are closer to the current experimental average~\cite{pdg2018}.
For the  $ h_c $ mode, both the current PQCD result and the previous calculations from~\cite{0607221,prd74114029}
are well consistent with the present data.
Two earlier papers~\cite{plb54271,prd69054023} studied  the decays  of $B^+ \rightarrow \chi_{c0} K^+$ and $B^+ \rightarrow h_c K^+$
by including the rescattering effects.
The evaluated branching ratios  lie in the ranges   $(1.1\sim 3.5)\times 10^{-4}$ and $(2\sim 12)\times 10^{-4}$, respectively.
Although the rescattering effects can enhance the branching ratio of  $B^+ \rightarrow \chi_{c0} K^+$ to match the data,
the value of $\mathcal {B}(B^+ \rightarrow h_c K^+)$  may be  overestimated owing to uncontrollable theoretical uncertainties.

From Table \ref{tab:br},
one can see  that the $B^+ \rightarrow \chi_{c2} K^+$ channel is predicted to have a threefold  larger  branching ratio
when compared with  the data,
which  is dominated by the Belle experiment~\cite{prl107091803}.
 $BABAR$ gave  the upper bounds (the values)~\cite{prl102132001}, as follows:
\begin{eqnarray}
\mathcal {B}(B^+ \rightarrow \chi_{c2} K^+)&<&1.8(1.0\pm0.6\pm0.1)\times 10^{-5},\nonumber\\
\mathcal {B}(B^0 \rightarrow \chi_{c2} K^0)&<&2.8(1.5\pm0.9\pm0.3)\times 10^{-5},
\end{eqnarray}
where the uncertainties are statistical and systematic, respectively.
It seems that their central values are somewhat different
and suffer from sizeable statistical uncertainties.
Our branching ratio for the neutral mode is comparable with  the  upper limit of  $BABAR$~\cite{prl102132001}.
It is worth noting that the predictions from QCDF on this mode have a relatively big spread.
For example, the QCDF prediction from~\cite{0607221} led to $\mathcal {B}(B^+ \rightarrow \chi_{c2} K^+)=3.0\times 10^{-6}$,
which is too small compared to the measured value.
However, another QCDF prediction, of the order of $10^{-4}$~\cite{plb619313}, is too large.
As pointed out in~\cite{plb619313}, the number can be adjusted to
the right magnitude with an appropriate choice for the parameters of the spectator hard scattering contributions.
The large numerical difference between  the two QCDF calculations is mainly caused by the large twist-3 spectator contribution,
which can be traced to the infrared behavior of the spectator interactions.

As mentioned above, the charged and neutral decay modes differ in the lifetimes of $B^0$ and $B^+$ in our formalism;
the predicted  branching ratios have almost  the same magnitude.
However, the data of $\mathcal {B}(B^0 \rightarrow \chi_{c0} K^0)$  in Table~\ref{tab:br}
is two orders of magnitude smaller than those of the  charged  one.
This number was obtained in the  LHCb experiment from the Dalitz plot (DP) analysis of the $B^0\rightarrow K_S^0\pi^+\pi^-$  decays~\cite{prl120261801}.
It indicates a tension with the similar amplitude analysis performed in the  $BABAR$ experiment~\cite{prd85112010,prd80112001,prd85054023}.
One can see that all   the model calculations in Table~\ref{tab:br}
 are substantially larger in magnitude than the LHCb  data.
Improved measurements are certainly needed for this decay mode.

The decays with  $\pi $ and $\bar K$ in the final state  have relatively small
branching ratios ($\sim 10^{-6}$) owing to the CKM factor suppression.
Experimentally, only the $BABAR$ collaboration reported the upper limits of the branching ratio products
$\mathcal {B}(B^+\rightarrow \chi_{c0,c2})\times \mathcal {B}(\chi_{c0,c2}\rightarrow \pi^+\pi^-) <1.0 \times 10^{-7}$
 by applying DP analysis on the  charmless decay $B\rightarrow \pi\pi\pi$.
Combining the experimental facts $\mathcal {B}(\chi_{c0}\rightarrow \pi^+\pi^-)=\frac{2}{3}(8.51\pm 0.33)\times 10^{-3}$
and $\mathcal {B}(\chi_{c2}\rightarrow \pi^+\pi^-)=\frac{2}{3}(2.23\pm 0.09)\times 10^{-3}$~\cite{pdg2018},
where the factors of $2/3$ are attributed to isospin,
we  can infer the  experimental upper bounds:
\begin{eqnarray}
\mathcal {B}_{\text{exp}}(B^+ \rightarrow \chi_{c0} \pi^+)&<&1.8\times 10^{-5},\nonumber\\
\mathcal {B}_{\text{exp}}(B^+ \rightarrow \chi_{c2} \pi^+)&<&6.7\times 10^{-5}.
\end{eqnarray}
There is substantial room for our predictions to reach the experimental upper limits.
As these  Cabibbo-suppressed decays have   received less attention in other approaches, we await future comparisons.

The $CP$ asymmetry arises from the interference between the tree and penguin amplitudes.
Using the same definition as that in Ref.~\cite{epjc78463},
we study direct $CP$ violation $A^{\text{dir}}_{CP}$  and mixing-induced
$CP$ asymmetry $S$ for  the decays in question.
The two $CP$-violating parameters   can be expressed as
 \begin{eqnarray}\label{eq:cps}
A^{\text{dir}}_{CP}=\frac{|\mathcal {\bar{A}}|^2-|\mathcal {A}|^2}{|\mathcal {\bar{A}}|^2+|\mathcal {A}|^2},\quad
S_{f}=\frac{2\text{Im}(\lambda_f)}{1+|\lambda_f|^2},
\end{eqnarray}
with $\lambda_f=\eta_f e^{-2i\beta_{(s)}}\frac{\mathcal {\bar{A}}}{\mathcal {A}}$
and $\eta_f=1(-1)$ for $CP$-even (odd) final  states.
$\beta_{(s)}$ is one of the angles of the unitarity triangle in the CKM quark-mixing matrix~\cite{pdg2018}.
Our numerical results are tabulated in Table~\ref{tab:cp},
where the errors are induced by the same sources as the ones expressed in Eq.~(\ref{eq:brs}).
Since the hadronic parameter  dependence canceled out in Eq.~(\ref{eq:cps}),
these $CP$ asymmetries are more sensitive to the heavy quark masses and the hard scale.
\begin{table}
\caption{PQCD predictions for  $CP$ asymmetry parameters.}
\label{tab:cp}
\begin{tabular}[t]{lcc}
\hline\hline
Modes  & $A^{\text{dir}}_{CP}$  & $S_{f}(\%)$    \\ \hline
$B^0 \rightarrow \chi_{c0} K_S^0 $  &$-5.5^{+0.1+0.0+0.4+3.3+2.5}_{-0.2-0.0-0.9-2.1-3.2}\times 10^{-4}$
&$-69.8^{+0.0+0.0+0.0+0.0+0.1}_{-0.0-0.0-0.0-0.0-0.1}$ \\
$B^0  \rightarrow \chi_{c2} K_S^0 $  &$-5.2^{+0.1+0.0+0.2+1.5+2.7}_{-0.0-0.0-0.1-0.0-3.0}\times 10^{-4}$
&$-69.8^{+0.0+0.0+0.0+0.0+0.1}_{-0.0-0.0-0.0-0.0-0.1}$ \\
$B^0  \rightarrow h_c K_S^0 $  &$1.3^{+0.0+0.0+0.0+0.2+1.0}_{-0.0-0.0-0.0-0.1-0.7}\times 10^{-4}$
&$69.4^{+0.0+0.0+0.0+0.0+0.0}_{-0.0-0.0-0.0-0.0-0.0}$  \\
$B^0  \rightarrow \chi_{c0} \pi^0 $  &$1.2^{+0.1+0.0+0.3+0.5+0.6}_{-0.0-0.0-0.2-0.7-0.4}\times 10^{-2}$
 &$-63.8^{+0.1+0.0+0.1+0.3+1.0}_{-0.2-0.0-0.2-0.4-0.2}$ \\
$B^0  \rightarrow \chi_{c2} \pi^0 $  &$9.9^{+0.3+0.0+0.5+0.1+6.1}_{-0.1-0.0-0.5-2.1-5.1}\times 10^{-3}$
&$-63.6^{+0.1+0.0+0.0+0.5+0.6}_{-0.1-0.0-0.0-0.4-0.6}$ \\
$B^0  \rightarrow h_c \pi^0 $  &$-2.5^{+0.1+0.0+0.1+0.6+1.6}_{-0.0-0.0-0.0-0.0-2.0}\times 10^{-3}$
&$70.8^{+0.1+0.0+0.0+0.1+0.2}_{-0.1-0.0-0.0-0.2-0.2}$ \\
$B_s^0  \rightarrow \chi_{c0} K_S^0$  &$1.1^{+0.1+0.0+0.3+0.7+0.6}_{-0.1-0.0-0.1-0.7-0.3}\times 10^{-2}$
&$4.5^{+0.2+0.0+0.2+0.4+0.7}_{-0.2-0.0-0.2-0.2-0.8}$ \\
$B_s^0  \rightarrow \chi_{c2} K_S^0 $  &$9.6^{+0.3+0.0+0.2+0.7+3.9}_{-0.7-0.0-0.6-2.8-4.4}\times 10^{-3}$
&$4.9^{+0.2+0.0+0.0+0.4+0.8}_{-0.2-0.0-0.0-0.5-0.6}$  \\
$B_s^0  \rightarrow h_c K_S^0 $        &$-2.5^{+0.0+0.0+0.1+0.8+1.3}_{-0.0-0.0-0.1-0.3-1.7}\times 10^{-3}$
 &$4.2^{+0.0+0.0+0.0+0.1+0.3}_{-0.1-0.0-0.0-0.2-0.3}$ \\
\hline\hline
\end{tabular}
\end{table}
It is evident that direct $CP$ violations for decays involving  $K$  are very small, of the order $10^{-4}$,
owing to an almost null weak phase from the CKM matrix element $V_{ts}$.
For other channels, the weak phase from $V_{td}$ will  enhance  $A^{\text{dir}}_{CP}$ substantially to the level of $10^{-2}$.
Because the penguin contribution is small compared to that of the tree,
the mixing-induced $CP$ asymmetry $S$ in Eq.~(\ref{eq:cps}) is approximately expressed as  $-\eta_f S\approx \sin(2\beta)$.
It is clear from Table~\ref{tab:cp} that the predicted values of $S$ are not significantly different from 
the current world average values
$\sin(2\beta)=0.695\pm0.019$ and $2\beta_s=(2.55\pm 1.15)\times 10^{-2}$~\cite{pdg2018}.
With further experimental data in the future, these modes can serve as alternatives to extract the CKM phases $\beta_{(s)}$.

Experimentally, the world average of $A_{CP}(B^+\rightarrow \chi_{c0}K^+)=-0.20\pm 0.18$ based on the measurements
$-0.14\pm0.15^{+0.03}_{-0.06}$~\cite{prd78012004} and  $-0.96\pm0.37\pm 0.04$~\cite{prd84092007} from $BABAR$
and  $-0.065\pm0.20^{+0.035}_{-0.024}$~\cite{prl96251803} from Belle.
It should be noted that the number from~\cite{prd84092007} has a large and non-Gaussian uncertainty
and its difference from zero is not statistically significant.
All measured direct $CP$ violations are consistent with zero.
Because LHCb has measured $A_{CP}$ to the accuracy of $10^{-3}$,
it is conceivable that an observation of $CP$ violation in other
decays  will be feasible in the near future.
The mixing-induced $CP$ asymmetry for $B^0\rightarrow\chi_{c0}K^0_S$ decay
is measured by $BABAR$~\cite{prd80112001} with two solutions:
\begin{eqnarray}
S(B^0\rightarrow \chi_{c0} K^0_S)&=&-0.69\pm0.52\pm0.04\pm0.07, \quad \text{Solution I}, \nonumber\\
S(B^0\rightarrow \chi_{c0} K^0_S)&=&-0.85\pm0.34\pm0.04\pm0.07, \quad \text{Solution II},
\end{eqnarray}
where the last uncertainty represents the dependence of the DP signal model.
We can note that the first experimental solution is more favored by our calculation.
\section{ conclusion}\label{sec:sum}

The factorization-forbidden decays of the  $B$ meson to charmonium
have been revisited in the PQCD formalism,
which is free from  endpoint singularities.
The charmonium distribution amplitudes are updated based on our previous study.
We find that the  dominant contribution comes from the leading-twist charmonium distribution amplitudes.
The constructive  interference between the two nonfactorizable spectator diagrams
enhances the decay rate, which can be compatible with the factorization-allowed decays.
The obtained branching ratios of the $B^+\rightarrow \chi_{c0} K^+$  and $B^+\rightarrow h_c K^+$ decays are
essentially in agreement with the current data,
whereas estimates of the $\chi_{c2}$ one is found to be larger, typically by a factor of 3.
For the  decays involving $\pi$  or $\bar K$ in the final state not yet measured,
the calculated branching ratios will be further tested by the LHCb and Belle-II experiments in the near future.
We further  estimate the $CP$-violating parameters.
As expected, the direct $CP$ asymmetries in these
channels are very low  owing to the suppressed penguin contributions.
The mixing-induced $CP$ asymmetries are close to $\sin(\beta_{(s)})$,
which suggests that these channels can provide a cross-check on the measurement
of the  CKM  phases $\beta$ and $\beta_s$.

We have also collected other theoretical results, whenever available, in Table~\ref{tab:br}
and made a detailed comparison.
The predicted branching ratios for most decay processes have similar magnitudes,
whereas they can differ by several factors for specific decay modes.
In general, our predictions are more consistent with the data compared to  these earlier analyses.

We discussed the theoretical uncertainties arising from
the hadronic parameters, such as $\omega$, $f_B$, and $m_0$, in the meson wave function,
heavy quark masses  and hard scale,
which can significantly  affect  the branching ratios,
whereas the $CP$ asymmetries
are found to be relatively stable with respect to variations in the  hadronic parameters.
The reasonably accurate results obtained with
will be tested at existing and forthcoming hadron colliders.
\begin{acknowledgments}
This work is supported in
part by the National Natural Science Foundation of China
under Grants No.11605060 and No.11547020,
by the Natural Science Foundation of Hebei Province under Grant No. A2019209449,
and by Department of Education of Hebei Province under Grant No. BJ2016041.
\end{acknowledgments}

\begin{appendix}
\section{Meson distribution amplitudes }\label{sec:app}
The light-cone  meson distribution amplitudes are, in principle, not calculable in PQCD,
but they are universal for all the decay channels.
In this Appendix we present  explicit expressions for
the meson distribution amplitudes appearing in the decay amplitudes in Section~\ref{sec:framework}.
First, for the $B$ meson distribution amplitudes, we adopt the model \cite{pqcd3,pqcd4,prd76074018}
\begin{eqnarray}
\phi_{B}(x,b)=N x^2(1-x)^2\exp[-\frac{x^2M^2}{2\omega^2_b}-\frac{\omega^2_bb^2}{2}],
\end{eqnarray}
with the shape parameters $\omega_b=0.4$ GeV and $\omega_b=0.5$ GeV for the $B$ and $B_s$ mesons, respectively.
The coefficient $N$ is related to the decay constant $f_B$ by normalization:
\begin{eqnarray}
\int_0^1\phi_{B}(x,b=0)d x=\frac{f_{B}}{2\sqrt{6}}.
\end{eqnarray}

The distribution amplitudes of the $P$-wave charmonium states, defined via the nonlocal matrix
element, have been derived in Ref. \cite{prd97033001};
we collect their expressions as following:
\begin{eqnarray}\label{eq:cDAs}
\psi^v_S(x)&=& \frac{f_S}{2\sqrt{6}}N_Tx(1-x)(2x-1)\mathcal {T}(x) \nonumber\\
\psi^s_S(x)&=& \frac{f_S}{2\sqrt{6}}N_S\mathcal {T}(x),\nonumber\\
\psi^L_A(x)&=& \frac{f_A}{2\sqrt{6}}N_Tx(1-x)(2x-1)\mathcal {T}(x),\nonumber\\
\psi^t_A(x)&=& \frac{f_A^{\perp}}{2\sqrt{6}}\frac{N_L}{2}(1-2x)^2\mathcal {T}(x),\nonumber\\
\psi_T(x)&=& \frac{f_T}{2\sqrt{6}}N_Tx(1-x)(2x-1)\mathcal {T}(x),\nonumber\\
\psi^t_T(x)&=& \frac{f_T^{\perp}}{2\sqrt{6}}\frac{N_T}{4}(2x-1)[1-6x+6x^2]\mathcal {T}(x),
\end{eqnarray}
where
 \begin{eqnarray}\label{eq:vv2}
\mathcal {T}(x)=\Big \{\frac{\sqrt{x(1-x)(1-4x(1-x))^3}}{[1-4x(1-x)(1-v^2/4)]^2}\Big \}^{1-v^2},
\end{eqnarray}
and  $v$ is the charm quark velocity,
which denote the relativistic corrections to the Coulomb wave functions \cite{plb612215}.
We consider $v^2=0.3$ for charmonium.
The normalization constants $N_{L, T, S}$ can be determined using the corresponding normalization conditions  \cite{prd97033001}.

The kaon and pion meson distribution amplitudes up to twist-3 are determined using the light-cone QCD sum rules \cite{jhep01010,npb529323}:
\begin{eqnarray}
\phi_K^A(x)&=&\frac{3f_K}{\sqrt{6}}x(1-x)[1+a_1^KC_1^{3/2}(2x-1)+a_2^KC_2^{3/2}(2x-1)],\nonumber\\
\phi_K^P(x)&=&\frac{f_K}{2\sqrt{6}}[1+0.24C_2^{1/2}(2x-1)],\nonumber\\
\phi_K^T(x)&=&\frac{f_K}{2\sqrt{6}}(1-2x)[1+0.35(10x^2-10x+1)],\nonumber\\
\phi_{\pi}^A(x)&=&\frac{3f_{\pi}}{\sqrt{6}}x(1-x)[1+a_2^{\pi}C_2^{3/2}(2x-1)],\nonumber\\
\phi_{\pi}^P(x)&=&\frac{f_{\pi}}{2\sqrt{6}}[1+0.43C_2^{1/2}(2x-1)],\nonumber\\
\phi_{\pi}^T(x)&=&\frac{f_{\pi}}{2\sqrt{6}}(1-2x)[1+0.55(10x^2-10x+1)],\nonumber\\
\end{eqnarray}
with the Gegenbauer polynomials
\begin{eqnarray}
C_1^{3/2}(x)=3x,\quad C_2^{3/2}(x)=1.5(5x^2-1),\quad C_2^{1/2}(x)=(3x^2-1)/2.
\end{eqnarray}
The Gegenbauer moments for the twist-2 LCDAs are used with the following updated values at the scale $\mu=1$ GeV \cite{jhep05004}:
\begin{eqnarray}
a_1^K=0.06, \quad a_2^{K/\pi}=0.25.
\end{eqnarray}

\end{appendix}

\end{document}